\documentclass[12pt,reqno]{amsart}
\usepackage{graphicx}
\usepackage{amscd}
\usepackage{amsmath}
\usepackage{epsfig}
\usepackage{amsfonts}
\usepackage{amssymb}

\providecommand{\U}[1]{\protect\rule{.1in}{.1in}}
\providecommand{\U}[1]{\protect\rule{.1in}{.1in}}
\allowdisplaybreaks

\theoremstyle{plain}

\newtheorem{lemma}{Lemma}

\numberwithin{equation}{section}

\input{tcilatex}

\begin{document}
\title[Harmonic Oscillator Group]{On The Harmonic Oscillator Group}
\author{Raquel M. L\'{o}pez}
\address{Mathematical, Computational and Modeling Sciences Center, Arizona
State University, Tempe, AZ 85287--1904, U.S.A.}
\email{rlopez14@asu.edu}
\author{Sergei K. Suslov}
\address{School of Mathematical and Statistical Sciences \& Mathematical,
Computational and Modeling Sciences Center, Arizona State University, Tempe,
AZ 85287--1804, U.S.A.}
\email{sks@asu.edu}
\urladdr{http://hahn.la.asu.edu/\symbol{126}suslov/index.html}
\author{Jos\'{e} M. Vega-Guzm\'{a}n}
\address{Mathematical, Computational and Modeling Sciences Center, Arizona
State University, Tempe, AZ 85287--1904, U.S.A.}
\email{jmvega@asu.edu}
\date{\today }

\begin{abstract}
We discuss the maximum kinematical invariance group of the quantum harmonic
oscillator from a viewpoint of the Ermakov-type system. A six parameter
family of the square integrable oscillator wave functions, which seems
cannot be obtained by the standard separation of variables, is presented as
an example. The invariance group of the generalized driven harmonic
oscillator is shown to be isomorphic to the corresponding Schr\"{o}dinger
group of the free particle.
\end{abstract}

\maketitle

Quantum systems with quadratic Hamiltonians are called the generalized
harmonic oscillators (see, for example, \cite{Cor-Sot:Sua:SusInv}, \cite%
{Dod:Mal:Man75}, \cite{Fey:Hib}, \cite{Malkin:Man'ko79}, \cite{Wolf81}, \cite%
{Yeon:Lee:Um:George:Pandey93}, \cite{Zhukov99} and the references therein).
These systems have attracted substantial attention over the years because of
their great importance in many advanced quantum problems. Examples are
coherent states and uncertainty relations, Berry's phase, quantization of
mechanical systems and Hamiltonian cosmology. More applications include, but
are not limited to charged particle traps and motion in uniform magnetic
fields, molecular spectroscopy and polyatomic molecules in varying external
fields, crystals through which an electron is passing and exciting the
oscillator modes, and other mode interactions with external fields.
Quadratic Hamiltonians have particular applications in quantum
electrodynamics because the electromagnetic field can be represented as a
set of forced harmonic oscillators \cite{Fey:Hib}.

The purpose of this paper is to give a simple derivation of the maximum
kinematical invariance groups of the free particle and harmonic oscillator,
which were introduced in Refs.~\cite{AndersonPlus72}, \cite{AndersonII72}, 
\cite{Hagen72}, \cite{Niederer72} and \cite{Niederer73} (see also \cite%
{BoySharpWint} and the references therein), from a unified approach to
generalized harmonic oscillators (see, for example, \cite%
{Cor-Sot:Lop:Sua:Sus}, \cite{Cor-Sot:Sua:SusInv}, \cite{Lan:Lop:Sus} and the
references therein). Relations with the corresponding Riccati and
Ermakov-type systems, which seem to be missing in the available literature,
are emphasized.

\section{Transforming Nonautonomous Schr\"{o}dinger Equation into Autonomous
Form}

Quantum systems described by the one-dimensional time-dependent Schr\"{o}%
dinger equation%
\begin{equation}
i\frac{\partial \psi }{\partial t}=H\psi ,  \label{Schroedinger}
\end{equation}%
where the variable Hamiltonian $H=Q\left( p,x\right) $ is an arbitrary
quadratic of two operators $p=-i\partial /\partial x$ and $x,$ namely,%
\begin{equation}
i\psi _{t}=-a\left( t\right) \psi _{xx}+b\left( t\right) x^{2}\psi -ic\left(
t\right) x\psi _{x}-id\left( t\right) \psi -f\left( t\right) x\psi +ig\left(
t\right) \psi _{x}  \label{SchroedingerQuadratic}
\end{equation}%
($a,$ $b,$ $c,$ $d,$ $f$ and $g$ are suitable real-valued functions of time
only), are known as the \textit{generalized (driven) harmonic oscillators}.
Some examples, a general approach and known elementary solutions can be
found in Refs.~\cite{Cor-Sot:Lop:Sua:Sus}, \cite{Cor-Sot:Sua:Sus}, \cite%
{Cor-Sot:Sua:SusInv}, \cite{Cor-Sot:Sus}, \cite{Dod:Mal:Man75}, \cite%
{Fey:Hib}, \cite{Lop:Sus}, \cite{Suaz:Sus}, \cite{Wolf81} and \cite%
{Yeon:Lee:Um:George:Pandey93}.

In this paper, we shall use the following result established in \cite%
{Lan:Lop:Sus}.

\begin{lemma}
The substitution%
\begin{equation}
\psi =\frac{e^{i\left( \alpha \left( t\right) x^{2}+\delta \left( t\right)
x+\kappa \left( t\right) \right) }}{\sqrt{\mu \left( t\right) }}\ \chi
\left( \xi ,\tau \right) ,\qquad \xi =\beta \left( t\right) x+\varepsilon
\left( t\right) ,\quad \tau =-\gamma \left( t\right)  \label{Ansatz}
\end{equation}%
transforms the non-autonomous and inhomogeneous Schr\"{o}dinger equation (%
\ref{SchroedingerQuadratic}) into the autonomous form%
\begin{equation}
i\chi _{\tau }=-\chi _{\xi \xi }+c_{0}\xi ^{2}\chi \qquad \left(
c_{0}=0,1\right)  \label{ASEq}
\end{equation}%
provided that%
\begin{equation}
\frac{d\alpha }{dt}+b+2c\alpha +4a\alpha ^{2}=c_{0}a\beta ^{4},  \label{SysA}
\end{equation}%
\begin{equation}
\frac{d\beta }{dt}+\left( c+4a\alpha \right) \beta =0,  \label{SysB}
\end{equation}%
\begin{equation}
\frac{d\gamma }{dt}+a\beta ^{2}=0  \label{SysC}
\end{equation}%
and%
\begin{equation}
\frac{d\delta }{dt}+\left( c+4a\alpha \right) \delta =f+2g\alpha
+2c_{0}a\beta ^{3}\varepsilon ,  \label{SysD}
\end{equation}%
\begin{equation}
\frac{d\varepsilon }{dt}=\left( g-2a\delta \right) \beta ,  \label{SysE}
\end{equation}%
\begin{equation}
\frac{d\kappa }{dt}=g\delta -a\delta ^{2}+c_{0}a\beta ^{2}\varepsilon ^{2}.
\label{SysF}
\end{equation}%
Here%
\begin{equation}
\alpha =\frac{1}{4a}\frac{\mu ^{\prime }}{\mu }-\frac{d}{2a}.  \label{Alpha}
\end{equation}
\end{lemma}

(We have changed $\tau \rightarrow -\tau $ in this paper for convenience. A 
\texttt{Mathematica} proof of Lemma~1 is given by Christoph Koutschan \cite%
{Kouchan11}; see also \cite{KouchanZeilberger10}. A special case of the
substitution (\ref{Ansatz}) has been named the Quantum Arnold Transformation
in the recent publications \cite{Ald:Coss:Guerr:Lop-Ru11}, \cite%
{Guerr:Lop:Ald:Coss11} and \cite{LopJRLR11}; see also \cite{Lan:Lop:Sus}, 
\cite{Zhukov99} and references therein for the earlier works.)

The substitution (\ref{Alpha}) reduces the inhomogeneous equation (\ref{SysA}%
) to the second order ordinary differential equation%
\begin{equation}
\mu ^{\prime \prime }-\tau \left( t\right) \mu ^{\prime }+4\sigma \left(
t\right) \mu =c_{0}\left( 2a\right) ^{2}\beta ^{4}\mu ,  \label{CharEq}
\end{equation}%
that has the familiar time-varying coefficients%
\begin{equation}
\tau \left( t\right) =\frac{a^{\prime }}{a}-2c+4d,\qquad \sigma \left(
t\right) =ab-cd+d^{2}+\frac{d}{2}\left( \frac{a^{\prime }}{a}-\frac{%
d^{\prime }}{d}\right) .  \label{TauSigma}
\end{equation}

When $c_{0}=0$, equation (\ref{SysA}) is called the \textit{Riccati
nonlinear differential equation} \cite{Wa}, \cite{Wh:Wa} and the system (\ref%
{SysA})--(\ref{SysF}) shall be referred to as a \textit{Riccati-type system}%
. (Similar terminology is used in \cite{SuazoSusVega10}, \cite%
{SuazoSusVega11} for the corresponding parabolic equation.) If $c_{0}=1,$\
equation (\ref{CharEq}) can be reduced to a generalized version of the 
\textit{Ermakov nonlinear differential equation} (see, for example, \cite%
{Cor-Sot:Sua:SusInv}, \cite{Ermakov}, \cite{Leach:Andrio08}, \cite{Suslov10}
and the references therein regarding Ermakov's equation) and we shall refer
to the corresponding system (\ref{SysA})--(\ref{SysF}) with $c_{0}\neq 0$ as
an \textit{Ermakov-type system}. Throughout this paper, we use the notations
from Ref.~\cite{Lan:Lop:Sus} where a more detailed bibliography on the
quadratic systems can be found. We have to remind to the reader how to solve
the systems (\ref{SysA})--(\ref{SysF}) (in quadratures) in order to make our
presentation as self-contained as possible.

The time-dependent coefficients $\alpha _{0},$ $\beta _{0},$ $\gamma _{0},$ $%
\delta _{0},$ $\varepsilon _{0},$ $\kappa _{0}$ that satisfy the
Riccati-type system (\ref{SysA})--(\ref{SysF}) are given as follows \cite%
{Cor-Sot:Lop:Sua:Sus}, \cite{Suaz:Sus}, \cite{Suslov10}:%
\begin{eqnarray}
&&\alpha _{0}\left( t\right) =\frac{1}{4a\left( t\right) }\frac{\mu
_{0}^{\prime }\left( t\right) }{\mu _{0}\left( t\right) }-\frac{d\left(
t\right) }{2a\left( t\right) },  \label{A0} \\
&&\beta _{0}\left( t\right) =-\frac{\lambda \left( t\right) }{\mu _{0}\left(
t\right) },\qquad \lambda \left( t\right) =\exp \left( -\int_{0}^{t}\left(
c\left( s\right) -2d\left( s\right) \right) \ ds\right) ,  \label{B0} \\
&&\gamma _{0}\left( t\right) =\frac{1}{2\mu _{1}\left( 0\right) }\frac{\mu
_{1}\left( t\right) }{\mu _{0}\left( t\right) }+\frac{d\left( 0\right) }{%
2a\left( 0\right) }  \label{C0}
\end{eqnarray}%
and%
\begin{equation}
\delta _{0}\left( t\right) =\frac{\lambda \left( t\right) }{\mu _{0}\left(
t\right) }\int_{0}^{t}\left[ \left( f\left( s\right) -\frac{d\left( s\right) 
}{a\left( s\right) }g\left( s\right) \right) \mu _{0}\left( s\right) +\frac{%
g\left( s\right) }{2a\left( s\right) }\mu _{0}^{\prime }\left( s\right) %
\right] \frac{ds}{\lambda \left( s\right) },  \label{D0}
\end{equation}%
\begin{eqnarray}
\varepsilon _{0}\left( t\right) &=&-\frac{2a\left( t\right) \lambda \left(
t\right) }{\mu _{0}^{\prime }\left( t\right) }\delta _{0}\left( t\right)
+8\int_{0}^{t}\frac{a\left( s\right) \sigma \left( s\right) \lambda \left(
s\right) }{\left( \mu _{0}^{\prime }\left( s\right) \right) ^{2}}\left( \mu
_{0}\left( s\right) \delta _{0}\left( s\right) \right) \ ds  \label{E0} \\
&&\quad +2\int_{0}^{t}\frac{a\left( s\right) \lambda \left( s\right) }{\mu
_{0}^{\prime }\left( s\right) }\left( f\left( s\right) -\frac{d\left(
s\right) }{a\left( s\right) }g\left( s\right) \right) \ ds,  \notag
\end{eqnarray}%
\begin{eqnarray}
\kappa _{0}\left( t\right) &=&\frac{a\left( t\right) \mu _{0}\left( t\right) 
}{\mu _{0}^{\prime }\left( t\right) }\delta _{0}^{2}\left( t\right)
-4\int_{0}^{t}\frac{a\left( s\right) \sigma \left( s\right) }{\left( \mu
_{0}^{\prime }\left( s\right) \right) ^{2}}\left( \mu _{0}\left( s\right)
\delta _{0}\left( s\right) \right) ^{2}\ ds  \label{F0} \\
&&\quad -2\int_{0}^{t}\frac{a\left( s\right) }{\mu _{0}^{\prime }\left(
s\right) }\left( \mu _{0}\left( s\right) \delta _{0}\left( s\right) \right)
\left( f\left( s\right) -\frac{d\left( s\right) }{a\left( s\right) }g\left(
s\right) \right) \ ds  \notag
\end{eqnarray}%
$(\delta _{0}\left( 0\right) =-\varepsilon _{0}\left( 0\right) =g\left(
0\right) /\left( 2a\left( 0\right) \right) $ and $\kappa _{0}\left( 0\right)
=0)$ provided that $\mu _{0}$ and $\mu _{1}$ are standard solutions of
equation (\ref{CharEq}) with $c_{0}=0$ corresponding to the initial
conditions $\mu _{0}\left( 0\right) =0,$ $\mu _{0}^{\prime }\left( 0\right)
=2a\left( 0\right) \neq 0$ and $\mu _{1}\left( 0\right) \neq 0,$ $\mu
_{1}^{\prime }\left( 0\right) =0.$ (Proofs of these facts are outlined in
Refs.~\cite{Cor-Sot:Lop:Sua:Sus}, \cite{Cor-Sot:SusDPO} and \cite{Suaz:Sus}.
Extensions to the nonlinear Schr\"{o}dinger equations are discussed in Refs.~%
\cite{AblowClark91}, \cite{Clark88}, \cite{SuazoSuslovSol} and \cite%
{Suslov11}.)

Solution of the Riccati-type system in terms of a nonlinear superposition
principle is considered in \cite{Suaz:Sus}.

\begin{lemma}
The solution of the Riccati-type system (\ref{SysA})--(\ref{SysF}) is given
by%
\begin{eqnarray}
&&\mu \left( t\right) =2\mu \left( 0\right) \mu _{0}\left( t\right) \left(
\alpha \left( 0\right) +\gamma _{0}\left( t\right) \right) ,  \label{MKernel}
\\
&&\alpha \left( t\right) =\alpha _{0}\left( t\right) -\frac{\beta
_{0}^{2}\left( t\right) }{4\left( \alpha \left( 0\right) +\gamma _{0}\left(
t\right) \right) },  \label{AKernel} \\
&&\beta \left( t\right) =-\frac{\beta \left( 0\right) \beta _{0}\left(
t\right) }{2\left( \alpha \left( 0\right) +\gamma _{0}\left( t\right)
\right) }=\frac{\beta \left( 0\right) \mu \left( 0\right) }{\mu \left(
t\right) }\lambda \left( t\right) ,  \label{BKernel} \\
&&\gamma \left( t\right) =\gamma \left( 0\right) -\frac{\beta ^{2}\left(
0\right) }{4\left( \alpha \left( 0\right) +\gamma _{0}\left( t\right)
\right) }  \label{CKernel}
\end{eqnarray}%
and%
\begin{eqnarray}
\delta \left( t\right) &=&\delta _{0}\left( t\right) -\frac{\beta _{0}\left(
t\right) \left( \delta \left( 0\right) +\varepsilon _{0}\left( t\right)
\right) }{2\left( \alpha \left( 0\right) +\gamma _{0}\left( t\right) \right) 
},  \label{DKernel} \\
\varepsilon \left( t\right) &=&\varepsilon \left( 0\right) -\frac{\beta
\left( 0\right) \left( \delta \left( 0\right) +\varepsilon _{0}\left(
t\right) \right) }{2\left( \alpha \left( 0\right) +\gamma _{0}\left(
t\right) \right) },  \label{EKernel} \\
\kappa \left( t\right) &=&\kappa \left( 0\right) +\kappa _{0}\left( t\right)
-\frac{\left( \delta \left( 0\right) +\varepsilon _{0}\left( t\right)
\right) ^{2}}{4\left( \alpha \left( 0\right) +\gamma _{0}\left( t\right)
\right) }  \label{FKernel}
\end{eqnarray}%
in terms of the fundamental solution (\ref{A0})--(\ref{F0}) subject to the
arbitrary (finite) initial data $\mu \left( 0\right) ,$ $\alpha \left(
0\right) ,$ $\beta \left( 0\right) \neq 0,$ $\gamma \left( 0\right) ,$ $%
\delta \left( 0\right) ,$ $\varepsilon \left( 0\right) ,$ $\kappa \left(
0\right) .$
\end{lemma}

The following extension allows to solve the Ermakov-type system \cite%
{Lan:Lop:Sus}.

\begin{lemma}
The solution of the Ermakov-type system (\ref{SysA})--(\ref{SysF}) when $%
c_{0}=1\left( \neq 0\right) $ is given by%
\begin{eqnarray}
&&\mu =\mu \left( 0\right) \mu _{0}\sqrt{\beta ^{4}\left( 0\right) +4\left(
\alpha \left( 0\right) +\gamma _{0}\right) ^{2}},  \label{MKernelOsc} \\
&&\alpha =\alpha _{0}-\beta _{0}^{2}\frac{\alpha \left( 0\right) +\gamma _{0}%
}{\beta ^{4}\left( 0\right) +4\left( \alpha \left( 0\right) +\gamma
_{0}\right) ^{2}},  \label{AKernelOsc} \\
&&\beta =-\frac{\beta \left( 0\right) \beta _{0}}{\sqrt{\beta ^{4}\left(
0\right) +4\left( \alpha \left( 0\right) +\gamma _{0}\right) ^{2}}}=\frac{%
\beta \left( 0\right) \mu \left( 0\right) }{\mu \left( t\right) }\lambda
\left( t\right) ,  \label{BKernelOsc} \\
&&\gamma =\gamma \left( 0\right) -\frac{1}{2}\arctan \frac{\beta ^{2}\left(
0\right) }{2\left( \alpha \left( 0\right) +\gamma _{0}\right) },\quad
a\left( 0\right) >0  \label{CKernelOsc}
\end{eqnarray}%
and%
\begin{eqnarray}
\delta &=&\delta _{0}-\beta _{0}\frac{\varepsilon \left( 0\right) \beta
^{3}\left( 0\right) +2\left( \alpha \left( 0\right) +\gamma _{0}\right)
\left( \delta \left( 0\right) +\varepsilon _{0}\right) }{\beta ^{4}\left(
0\right) +4\left( \alpha \left( 0\right) +\gamma _{0}\right) ^{2}},
\label{DKernelOsc} \\
\varepsilon &=&\frac{2\varepsilon \left( 0\right) \left( \alpha \left(
0\right) +\gamma _{0}\right) -\beta \left( 0\right) \left( \delta \left(
0\right) +\varepsilon _{0}\right) }{\sqrt{\beta ^{4}\left( 0\right) +4\left(
\alpha \left( 0\right) +\gamma _{0}\right) ^{2}}},  \label{EKernelOsc} \\
\kappa &=&\kappa \left( 0\right) +\kappa _{0}-\varepsilon \left( 0\right)
\beta ^{3}\left( 0\right) \frac{\delta \left( 0\right) +\varepsilon _{0}}{%
\beta ^{4}\left( 0\right) +4\left( \alpha \left( 0\right) +\gamma
_{0}\right) ^{2}}  \label{FKernelOsc} \\
&&+\left( \alpha \left( 0\right) +\gamma _{0}\right) \frac{\varepsilon
^{2}\left( 0\right) \beta ^{2}\left( 0\right) -\left( \delta \left( 0\right)
+\varepsilon _{0}\right) ^{2}}{\beta ^{4}\left( 0\right) +4\left( \alpha
\left( 0\right) +\gamma _{0}\right) ^{2}}  \notag
\end{eqnarray}%
in terms of the fundamental solution (\ref{A0})--(\ref{F0}) subject to the
arbitrary initial data $\mu \left( 0\right) ,$ $\alpha \left( 0\right) ,$ $%
\beta \left( 0\right) \neq 0,$ $\gamma \left( 0\right) ,$ $\delta \left(
0\right) ,$ $\varepsilon \left( 0\right) ,$ $\kappa \left( 0\right) .$
\end{lemma}

Using standard oscillator wave functions for equation (\ref{ASEq}) when $%
c_{0}=1$ results in%
\begin{equation}
\psi _{n}\left( x,t\right) =\frac{e^{i\left( \alpha x^{2}+\delta x+\kappa
\right) +i\left( 2n+1\right) \gamma }}{\sqrt{2^{n}n!\mu \sqrt{\pi }}}\
e^{-\left( \beta x+\varepsilon \right) ^{2}/2}\ H_{n}\left( \beta
x+\varepsilon \right) ,  \label{WaveFunction}
\end{equation}%
where $H_{n}\left( x\right) $ are the Hermite polynomials \cite{Ni:Su:Uv}
and the solution of the Ermakov-type system (\ref{SysA})--(\ref{SysF}) is
available \cite{Lan:Lop:Sus}. They are also eigenfunctions,%
\begin{equation}
E\left( t\right) \psi _{n}\left( x,t\right) =\lambda \left( t\right) \left(
n+\frac{1}{2}\right) \psi _{n}\left( x,t\right) ,  \label{EigenValueProblem}
\end{equation}%
of the corresponding dynamic invariant ~\cite{SanSusVin}:%
\begin{equation}
E\left( t\right) =\frac{\lambda \left( t\right) }{2}\left[ \frac{\left(
p-2\alpha x-\delta \right) ^{2}}{\beta ^{2}}+\left( \beta x+\varepsilon
\right) ^{2}\right] ,\qquad \frac{d}{dt}\langle E\rangle =0.
\label{QuadraticInvariant}
\end{equation}%
The Green function of generalized harmonic oscillators,%
\begin{equation}
G\left( x,y,t\right) =\frac{1}{\sqrt{2\pi i\mu _{0}\left( t\right) }}\exp %
\left[ i\left( \alpha _{0}\left( t\right) x^{2}+\beta _{0}\left( t\right)
xy+\gamma _{0}\left( t\right) y^{2}+\delta _{0}\left( t\right) x+\varepsilon
_{0}\left( t\right) y+\kappa _{0}\left( t\right) \right) \right] ,
\label{GreenFunction}
\end{equation}%
has been constructed in Ref.~\cite{Cor-Sot:Lop:Sua:Sus}. Evaluation of the
Berry phase is discussed in Ref.~\cite{SanSusVin}.

In the reminder of the paper, we apply these general results to the maximum
kinematical invariance groups of the free Schr\"{o}dinger equation and of
the harmonic oscillator \cite{Niederer72}, \cite{Niederer73} (see also \cite%
{BoySharpWint} and the references therein). In addition, our approach allows
to describe the maximal kinematical invariance group of the generalized
harmonic oscillators.

\section{Special Cases}

In this section, we show that the maximal kinematical invariance groups of
the free Schr\"{o}dinger equation \cite{Niederer72} and the harmonic
oscillator \cite{Niederer73} (and their isomorphism) can be obtain as
special cases of the transformation (\ref{Ansatz}).

\subsection{Transformation from the free particle to the free particle}

In the simplest case $a=1,$ $b=c=d=f=g=c_{0}=0,$ one finds $\mu _{0}=2t,$ $%
\mu _{1}=1$ and $\alpha _{0}=-\beta _{0}/2=\gamma _{0}=1/\left( 4t\right) ,$ 
$\delta _{0}=\varepsilon _{0}=\kappa _{0}=0.$ The general solution of the
corresponding Riccati-type system is given by%
\begin{eqnarray}
&&\mu \left( t\right) =\mu \left( 0\right) \left( 1+4\alpha \left( 0\right)
t\right) , \\
&&\alpha \left( t\right) =\frac{\alpha \left( 0\right) }{1+4\alpha \left(
0\right) t},\qquad \qquad \beta \left( t\right) =\frac{\beta \left( 0\right) 
}{1+4\alpha \left( 0\right) t}, \\
&&\gamma \left( t\right) =\gamma \left( 0\right) -\frac{\beta ^{2}\left(
0\right) t}{1+4\alpha \left( 0\right) t},\quad \delta \left( t\right) =\frac{%
\delta \left( 0\right) }{1+4\alpha \left( 0\right) t}, \\
&&\varepsilon \left( t\right) =\varepsilon \left( 0\right) -\frac{2\beta
\left( 0\right) \delta \left( 0\right) t}{1+4\alpha \left( 0\right) t},\quad
\kappa \left( t\right) =\kappa \left( 0\right) -\frac{\delta ^{2}\left(
0\right) t}{1+4\alpha \left( 0\right) t}.
\end{eqnarray}%
The Ansatz (\ref{Ansatz}) together with these formulas determine the Schr%
\"{o}dinger group, namely, the maximum (known) kinematical invariance group
of the free Schr\"{o}dinger equation, as follows \cite{Niederer72}:%
\begin{eqnarray}
\psi \left( x,t\right) &=&\frac{1}{\sqrt{\mu \left( 0\right) \left(
1+4\alpha \left( 0\right) t\right) }}\ \exp \left[ i\left( \frac{\alpha
\left( 0\right) x^{2}+\delta \left( 0\right) x-\delta ^{2}\left( 0\right) t}{%
1+4\alpha \left( 0\right) t}+\kappa \left( 0\right) \right) \right]  \notag
\\
&&\times \chi \left( \frac{\beta \left( 0\right) x-2\beta \left( 0\right)
\delta \left( 0\right) t}{1+4\alpha \left( 0\right) t}+\varepsilon \left(
0\right) ,\ \frac{\beta ^{2}\left( 0\right) t}{1+4\alpha \left( 0\right) t}%
-\gamma \left( 0\right) \right) .  \label{SchroedingerGroup}
\end{eqnarray}%
We have established a connection of the Schr\"{o}dinger group with the
Riccati-type system (see also \cite{BoySharpWint}, \cite{GagWint93}, \cite%
{KalninsMiller74}, \cite{Miller77}, \cite{Niederer73}, \cite%
{VinetZhedanov2011} for the Lie group approach; subgroups of the Schr\"{o}%
dinger group and their invariants are discussed in \cite{BoySharpWint}).

The subgroups include the familiar \textit{Galilei transformations}:%
\begin{equation}
\psi \left( x,t\right) =\exp \left[ i\left( \frac{V}{2}x-\frac{V^{2}}{4}%
t\right) \right] \chi \left( x-Vt+x_{0},t-t_{0}\right) ,  \label{Galilei}
\end{equation}%
when $\alpha \left( 0\right) =\kappa \left( 0\right) =0,$ $\beta \left(
0\right) =\mu \left( 0\right) =1,$ $\gamma \left( 0\right) =t_{0},$ $%
\varepsilon \left( 0\right) =x_{0}$ and $\delta \left( 0\right) =V/2;$
supplemented by \textit{dilatations}:%
\begin{equation}
\psi \left( x,t\right) =\chi \left( lx,l^{2}t\right)  \label{dilatations}
\end{equation}%
with $\alpha \left( 0\right) =\gamma \left( 0\right) =\delta \left( 0\right)
=\varepsilon \left( 0\right) =\kappa \left( 0\right) =0,$ $\mu \left(
0\right) =1$ and $\beta \left( 0\right) =l;$ and \textit{expansions}:%
\begin{eqnarray}
\psi \left( x,t\right) &=&\frac{1}{\sqrt{1+mt}}\exp \left( i\frac{mx^{2}}{%
4\left( 1+mt\right) }\right) \chi \left( \frac{x}{1+mt},\ \frac{t}{1+mt}%
\right)  \label{expansions} \\
&&\left( \mu \left( 0\right) =1\left( \neq 0\right) ,\qquad \mu ^{\prime
}\left( 0\right) =m\right) ,  \notag \\
\psi \left( x,t\right) &=&\frac{1}{\sqrt{2t}}\exp \left( i\frac{x^{2}}{4t}%
\right) \chi \left( -\frac{x}{2t},\ -\frac{1}{4t}\right)
\label{expansiongreen} \\
&&\left( \mu \left( 0\right) =0,\qquad \mu ^{\prime }\left( 0\right)
=2\left( \neq 0\right) \right)  \notag
\end{eqnarray}%
with $\beta \left( 0\right) =1,$ $\delta \left( 0\right) =\varepsilon \left(
0\right) =\kappa \left( 0\right) =0$ $.$ (The symmetry group of the
corresponding diffusion equation is discussed in \cite{KalninsMiller74}, 
\cite{Miller77}, \cite{Rosen76} and \cite{SuazoSusVega11}.)

Here, we derive the Schr\"{o}dinger group for the free particle as a very
special case of Lemma~1. In turn, application of (\ref{SchroedingerGroup})
in (\ref{Ansatz}) produces a composition of these two transformations.

\subsection{Transformations from the harmonic oscillator to the free particle%
}

If $a=b=1,$ $c=d=f=g=c_{0}=0,$ the corresponding characteristic equation, $%
\mu ^{\prime \prime }+4\mu =0,$ has two standard solutions $\mu _{0}=\sin
2t, $ $\mu _{1}=\cos 2t$ and%
\begin{eqnarray}
&&\alpha _{0}=\gamma _{0}=\frac{\cos 2t}{2\sin 2t},\quad \beta _{0}=-\frac{1%
}{\sin 2t},\quad \delta _{0}=\varepsilon _{0}=\kappa _{0}=0, \\
&&\ \ \mu =\mu \left( 0\right) \left( 2\alpha \left( 0\right) \sin 2t+\cos
2t\right) .
\end{eqnarray}%
The general solution of the corresponding Riccati-type system takes the form%
\begin{eqnarray}
\alpha \left( t\right) &=&\frac{2\alpha \left( 0\right) \cos 2t-\sin 2t}{%
2\left( 2\alpha \left( 0\right) \sin 2t+\cos 2t\right) }, \\
\beta \left( t\right) &=&\frac{\beta \left( 0\right) }{2\alpha \left(
0\right) \sin 2t+\cos 2t}, \\
\gamma \left( t\right) &=&\gamma \left( 0\right) -\frac{\beta ^{2}\left(
0\right) \sin 2t}{2\left( 2\alpha \left( 0\right) \sin 2t+\cos 2t\right) },
\\
\delta \left( t\right) &=&\frac{\delta \left( 0\right) }{2\alpha \left(
0\right) \sin 2t+\cos 2t}, \\
\varepsilon \left( t\right) &=&\varepsilon \left( 0\right) -\frac{\beta
\left( 0\right) \delta \left( 0\right) \sin 2t}{2\alpha \left( 0\right) \sin
2t+\cos 2t}, \\
\kappa \left( t\right) &=&\kappa \left( 0\right) -\frac{\delta ^{2}\left(
0\right) \sin 2t}{2\left( 2\alpha \left( 0\right) \sin 2t+\cos 2t\right) }
\end{eqnarray}%
subject to arbitrary given initial data. Letting $\mu \left( 0\right) =\beta
\left( 0\right) =1$ and $\alpha \left( 0\right) =\gamma \left( 0\right)
=\delta \left( 0\right) =\varepsilon \left( 0\right) =\kappa \left( 0\right)
=0,$ we arrive at the simple substitution \cite{Niederer73}:%
\begin{equation}
\psi \left( x,t\right) =\frac{e^{-\left( i/2\right) x^{2}\tan 2t}}{\sqrt{%
\cos 2t}}\ \chi \left( \frac{x}{\cos 2t},\frac{\tan 2t}{2}\right)
\label{ClarkWint}
\end{equation}%
(see also \cite{GagWint93}, \cite{Guerr:Lop:Ald:Coss11}, \cite{Miller77}, 
\cite{Suslov11}).

\subsection{Transformation from the free particle to the harmonic oscillator}

In the simplest case $a=c_{0}=1,$ $b=c=d=f=g=0,$ the general solution of the
corresponding Ermakov-type system is given by%
\begin{eqnarray}
\mu \left( t\right) &=&\mu \left( 0\right) \sqrt{4\beta ^{4}\left( 0\right)
t^{2}+\left( 4\alpha \left( 0\right) t+1\right) ^{2}},  \label{fhM} \\
\alpha \left( t\right) &=&\frac{\beta ^{4}\left( 0\right) t+\alpha \left(
0\right) \left( 4\alpha \left( 0\right) t+1\right) }{4\beta ^{4}\left(
0\right) t^{2}+\left( 4\alpha \left( 0\right) t+1\right) ^{2}},  \label{fhA}
\\
\beta \left( t\right) &=&\frac{\beta \left( 0\right) }{\sqrt{4\beta
^{4}\left( 0\right) t^{2}+\left( 4\alpha \left( 0\right) t+1\right) ^{2}}},
\label{fhB} \\
\gamma \left( t\right) &=&\gamma \left( 0\right) -\frac{1}{2}\arctan \frac{%
2\beta ^{2}\left( 0\right) t}{4\alpha \left( 0\right) t+1},  \label{fhG} \\
\delta \left( t\right) &=&\frac{2\varepsilon \left( 0\right) \beta
^{3}\left( 0\right) t+\delta \left( 0\right) \left( 4\alpha \left( 0\right)
t+1\right) }{4\beta ^{4}\left( 0\right) t^{2}+\left( 4\alpha \left( 0\right)
t+1\right) ^{2}},  \label{fhD} \\
\varepsilon \left( t\right) &=&\frac{\varepsilon \left( 0\right) \left(
4\alpha \left( 0\right) t+1\right) -2\beta \left( 0\right) \delta \left(
0\right) t}{\sqrt{4\beta ^{4}\left( 0\right) t^{2}+\left( 4\alpha \left(
0\right) t+1\right) ^{2}}},  \label{fhE} \\
\kappa \left( t\right) &=&\kappa \left( 0\right) +t\ \frac{\left( 4\alpha
\left( 0\right) t+1\right) \left( \varepsilon ^{2}\left( 0\right) \beta
^{2}\left( 0\right) -\delta ^{2}\left( 0\right) \right) }{4\beta ^{4}\left(
0\right) t^{2}+\left( 4\alpha \left( 0\right) t+1\right) ^{2}}  \label{fhK}
\\
&&-t^{2}\ \frac{4\varepsilon \left( 0\right) \delta \left( 0\right) \beta
^{3}\left( 0\right) }{4\beta ^{4}\left( 0\right) t^{2}+\left( 4\alpha \left(
0\right) t+1\right) ^{2}}.  \notag
\end{eqnarray}%
When $\mu \left( 0\right) =\beta \left( 0\right) =1,$ $\alpha \left(
0\right) =\gamma \left( 0\right) =\delta \left( 0\right) =\varepsilon \left(
0\right) =\kappa \left( 0\right) =0,$ we get the known transformation \cite%
{Guerr:Lop:Ald:Coss11}, \cite{Hagen72}, \cite{JACKIW80}, \cite{LopJRLR11}, 
\cite{Miller77}, \cite{Niederer73}:%
\begin{equation}
\psi \left( x,t\right) =\frac{1}{\left( 4t^{2}+1\right) ^{1/4}}\exp \left( i%
\frac{tx^{2}}{4t^{2}+1}\right) \ \chi \left( \frac{x}{\sqrt{4t^{2}+1}},\frac{%
1}{2}\arctan 2t\right) .  \label{freeharmonic}
\end{equation}%
One can easily verify that transformations (\ref{ClarkWint}) and (\ref%
{freeharmonic}) are (local) inverses of each other.

Equations (\ref{WaveFunction}) and (\ref{fhM})--(\ref{fhK}) provide a six
parameter family of the \textquotedblleft harmonic oscillator
states\textquotedblright\ for the free particle (see also \cite%
{Guerr:Lop:Ald:Coss11} and the references therein).

\subsection{Transformation from the harmonic oscillator to the harmonic
oscillator}

We consider the case $a=b=c_{0}=1,$ $c=d=f=g=0.$ The general solution of the
corresponding Ermakov-type system is given by%
\begin{eqnarray}
\mu \left( t\right) &=&\mu \left( 0\right) \sqrt{\beta ^{4}\left( 0\right)
\sin ^{2}2t+\left( 2\alpha \left( 0\right) \sin 2t+\cos 2t\right) ^{2}},
\label{hhM} \\
\alpha \left( t\right) &=&\frac{\alpha \left( 0\right) \cos 4t+\sin 4t\
\left( \beta ^{4}\left( 0\right) +4\alpha ^{2}\left( 0\right) -1\right) /4}{%
\beta ^{4}\left( 0\right) \sin ^{2}2t+\left( 2\alpha \left( 0\right) \sin
2t+\cos 2t\right) ^{2}},  \label{hhA} \\
\beta \left( t\right) &=&\frac{\beta \left( 0\right) }{\sqrt{\beta
^{4}\left( 0\right) \sin ^{2}2t+\left( 2\alpha \left( 0\right) \sin 2t+\cos
2t\right) ^{2}}},  \label{hhB} \\
\gamma \left( t\right) &=&\gamma \left( 0\right) -\frac{1}{2}\arctan \frac{%
\beta ^{2}\left( 0\right) \sin 2t}{2\alpha \left( 0\right) \sin 2t+\cos 2t},
\label{hhG} \\
\delta \left( t\right) &=&\frac{\delta \left( 0\right) \left( 2\alpha \left(
0\right) \sin 2t+\cos 2t\right) +\varepsilon \left( 0\right) \beta
^{3}\left( 0\right) \sin 2t}{\beta ^{4}\left( 0\right) \sin ^{2}2t+\left(
2\alpha \left( 0\right) \sin 2t+\cos 2t\right) ^{2}},  \label{hhD} \\
\varepsilon \left( t\right) &=&\frac{\varepsilon \left( 0\right) \left(
2\alpha \left( 0\right) \sin 2t+\cos 2t\right) -\beta \left( 0\right) \delta
\left( 0\right) \sin 2t}{\sqrt{\beta ^{4}\left( 0\right) \sin ^{2}2t+\left(
2\alpha \left( 0\right) \sin 2t+\cos 2t\right) ^{2}}},  \label{hhE} \\
\kappa \left( t\right) &=&\kappa \left( 0\right) +\sin ^{2}2t\ \frac{%
\varepsilon \left( 0\right) \beta ^{2}\left( 0\right) \left( \alpha \left(
0\right) \varepsilon \left( 0\right) -\beta \left( 0\right) \delta \left(
0\right) \right) -\alpha \left( 0\right) \delta ^{2}\left( 0\right) }{\beta
^{4}\left( 0\right) \sin ^{2}2t+\left( 2\alpha \left( 0\right) \sin 2t+\cos
2t\right) ^{2}}  \label{hhK} \\
&&+\frac{1}{4}\sin 4t\ \frac{\varepsilon ^{2}\left( 0\right) \beta
^{2}\left( 0\right) -\delta ^{2}\left( 0\right) }{\beta ^{4}\left( 0\right)
\sin ^{2}2t+\left( 2\alpha \left( 0\right) \sin 2t+\cos 2t\right) ^{2}}. 
\notag
\end{eqnarray}%
(A direct substitution with the help of \texttt{Mathematica} verifies that
these expressions indeed satisfy the Ermakov-type system.)

The Ansatz (\ref{Ansatz}) together with the formulas (\ref{hhM})--(\ref{hhK}%
) explicitly determine the harmonic oscillator group introduced in Ref.~\cite%
{Niederer73}. The established connection with the corresponding Ermakov-type
system allows us to bypass the traditional Lie algebra approach \cite%
{AndersonPlus72}, \cite{BoySharpWint}, \cite{Miller77}, \cite{Niederer73}.
Combination of the transformations (\ref{SchroedingerGroup}) and (\ref%
{expansiongreen}), (\ref{ClarkWint}) and (\ref{freeharmonic}) implies that
the harmonic oscillator group is isomorphic to the Schr\"{o}dinger group of
the free particle (see (\ref{similarity}) below) \cite{Niederer73}. An
analog of the similarity transformation (\ref{expansiongreen}) takes the
form $\psi \left( x,t\right) =\chi \left( -x,t-\pi /4\right) $ and further
identification of the corresponding subgroups and the identity element is
similar to the case of the free particle. We have derived the harmonic
oscillator group as a special case of Lemma~1 and one can consider\ a
composition of these transformations.

Use of the explicit solution (\ref{hhM})--(\ref{hhK}) in the wave function (%
\ref{WaveFunction}) results in a six parameter (square integrable) family of
the \textquotedblleft dynamic quantum oscillator states\textquotedblright ,
which seems cannot be obtained by the standard separation of variables (the
case $\beta \left( 0\right) =1$ and $\alpha \left( 0\right) =\gamma \left(
0\right) =\delta \left( 0\right) =\varepsilon \left( 0\right) =\kappa \left(
0\right) =0$ corresponds to the textbook oscillator solution \cite{La:Lif}, 
\cite{Merz}). The corresponding quadratic dynamic invariant (\ref%
{QuadraticInvariant}) and the creation and annihilation operators are found
(in general) in Ref.~\cite{SanSusVin}. A generalization of the coherent
states is discussed in Ref.~\cite{Lan:Sus12}.

\section{The Maximal Kinematical Invariance Group of the Generalized Driven
Harmonic Oscillators}

The transformation (\ref{Ansatz}) from Lemma~1 admits an inversion when the
coefficient $a\left( t\right) $ does not change its sign (see (\ref{SysC})
for the monotonicity and local time inversion). As a result, the invariance
group of the generalized driven harmonic oscillator is isomorphic to the Schr%
\"{o}dinger group of the free particle,%
\begin{equation}
T=S^{-1}T_{0}S,  \label{similarity}
\end{equation}%
thus extending the result of \cite{Niederer73} to the corresponding
nonautonomous systems (in the classical case, see (\ref{ClarkWint}) and (\ref%
{freeharmonic}) for possible operators $S$ and $S^{-1}$ and the operator $%
T_{0}$ is defined by (\ref{SchroedingerGroup}) and (\ref{expansiongreen})).
The structure of the Schr\"{o}dinger group of operators $T_{0}$ in
two-dimensional space-time as a semidirect product of $SL\left( 2,%
\mathbb{R}
\right) $ and Weyl $W\left( 1\right) $ groups is discussed, for example, in
Refs.~\cite{BoySharpWint}, \cite{KalninsMiller74} and \cite{Miller77}.
Further details are left to the reader.

\section{A Conclusion}

Our analysis of the maximal kinematical\ invariance group of the quantum
harmonic oscillator provides a six parameter family of solutions, namely (%
\ref{WaveFunction}) and (\ref{hhM})--(\ref{hhK}), for the arbitrary initial
data of the corresponding Ermakov-type system. These \textquotedblleft
hidden parameters\textquotedblright\ disappear after evaluation of matrix
elements and cannot be observed from the spectrum. How to distinguish
between these \textquotedblleft new dynamic\textquotedblright\ and the
\textquotedblleft standard\textquotedblright\ harmonic oscillator states is
thus an open problem. (The probability density $\left\vert \psi \right\vert
^{2}$ of the solution (\ref{WaveFunction}) is moving with time even for the
oscillator \textquotedblleft dynamic ground state\textquotedblright\ -- a
simple \texttt{Mathematica} animation reveals such oscillations. This
effect, quite possibly, can be observed experimentally.) This ambiguity,
which is due to the nontrivial oscillator maximal kinematical\ invariance
group, then goes further into the coherent states, evaluation of Berry's
phase and dynamic invariants through the established connection with
solutions of the Ermakov-type system. All of that puts considerations of
this paper into a somewhat broader mathematical and physical context.

\noindent \textbf{Acknowledgments.\/} We thank Professor Carlos Castillo-Ch%
\'{a}vez, Professor Victor V. Dodonov, Dr. Christoph Koutschan, Professor
Vladimir I. Man'ko, Professor Peter Paule, Dr. Andreas Ruffing, Professor
Luc Vinet and Professor Doron Zeilberger for support, valuable discussions
and encouragement. The authors are indebted to Dr.~Francisco F.~L\'{o}%
pez-Ruiz for kindly pointing out the papers \cite{Ald:Coss:Guerr:Lop-Ru11}, 
\cite{Guerr:Lop:Ald:Coss11} and \cite{LopJRLR11} to our attention. This
paper has been initiated during a short visit of one of the authors (SKS) to
The Erwin Schr\"{o}dinger International Institute for Mathematical Physics
and we thank Professor Christian Krattenthaler, Fakult\"{a}t f\"{u}r
Mathematik, Universit\"{a}t Wien, for his hospitality. This research is
supported in part by the National Science Foundation--Enhancing the
Mathematical Sciences Workforce in the 21st Century (EMSW21), award \#
0838705; the Alfred P. Sloan Foundation--Sloan National Pipeline Program in
the Mathematical and Statistical Sciences, award \# LTR 05/19/09; and the
National Security Agency--Mathematical \& Theoretical Biology
Institute---Research program for Undergraduates; award \# H98230-09-1-0104.

\end{document}